\documentclass[aps,10pt,prc,showpacs,twocolumn,floatfix,nofootinbib]{revtex4}
\usepackage{graphicx}
\usepackage{amssymb}
\usepackage{amsmath}
\usepackage{bm}
\usepackage{color}
\newcommand{\Eq}[1]{Eq.\@ (\ref{#1})}
\newcommand{\Eqs}[1]{Eqs.\@ (\ref{#1})}
\newcommand{\Fig}[1]{Fig.\@ \ref{#1}}
\newcommand{\Figs}[1]{Figs.\@ \ref{#1}}
\newcommand{\Ref}[1]{Ref.\@ \cite{#1}}
\newcommand{\Refs}[1]{Refs.\@ \cite{#1}}
\newcommand{\Sec}[1]{Sec.\@ \ref{#1}}
\newcommand{\Secs}[1]{Secs.\@ \ref{#1}}
\newcommand{\vek}[1]{\bm{\mathrm{#1}}}
\newcommand{\nablav}{\vek{\nabla}}
\newcommand{\jv}{\vek{j}}
\newcommand{\pv}{\vek{p}}
\newcommand{\rv}{\vek{r}}
\newcommand{\sv}{\vek{s}}
\newcommand{\vv}{\vek{v}}
\newcommand{\calN}{\mathcal{N}}
\begin{document}
\title{Pygmy resonance and torus mode within Vlasov dynamics}
\author{Michael Urban}
\affiliation{Institut de Physique Nucl\'eaire, CNRS-IN2P3 and Universit\'e
Paris-Sud 11, 91406 Orsay cedex, France}
\begin{abstract}
The pygmy dipole resonance in neutron-rich nuclei is studied within
the framework of the Vlasov equation which is solved numerically. The
interaction used in the Thomas-Fermi ground state and in the Vlasov
equation is derived from an energy functional which correctly
describes the equation of state of nuclear matter and neutron
matter. It is found that the pygmy resonance appears in the electric
dipole response of all nuclei with strong neutron excess, the energies
and transition probabilities being in reasonable agreement with
experimental results. Since the Vlasov equation does not account for
any shell effects, this indicates that the existence of the pygmy
resonance is a generic phenomenon and does not rely on the specific
shell structure. Besides the electric dipole response, the isoscalar
toroidal response is calculated. The transition densities and velocity
fields are discussed. A comparison of the peak positions and velocity
fields suggests that the pygmy resonance can be identified with one of
the low-lying modes excited by the isoscalar toroidal operator.
\end{abstract}
\pacs{21.10.Re, 24.30.Cz, 03.65.Sq, 02.70.Ns}
\maketitle
\section{Introduction}
\label{subsec:introduction}
Neutron rich nuclei have become a very popular object of experimental
and theoretical nuclear structure studies. Besides the crucial role
these nuclei play in nuclear astrophysics and their importance for
constraining the nuclear energy density functional, these nuclei
exhibit fascinating properties which are qualitatively different from
ordinary nuclei. For instance, as a consequence of the ``neutron
skin'' surrounding the core of medium-mass and heavy neutron-rich
nuclei, there are new kinds of collective motion which are absent in
nuclei without strong neutron excess (see \cite{Paar} for a recent
review).

A famous example for such a collective mode is the so-called pygmy
resonance. Contrary to the well-known isovector giant-dipole resonance
(GDR), where neutrons and protons move against each other, the
pygmy-dipole resonance (PDR) consists, roughly speaking, of an
oscillation of the neutron skin against the $N=Z$ core. This mode is
not only interesting in itself, but its existence has also a strong
effect on the abundances of the elements in the universe
\cite{Goriely}. After first studies within schematic hydrodynamic
models \cite{Mohan,Suzuki}, the pygmy mode was investigated within the
random-phase approximation (RPA), using non-relativistic \cite{Catara}
or relativistic formalisms \cite{VretenarNPA,Pena}, and beyond, using
the quasiparticle-phonon model \cite{Ryezayeva,Tsoneva}.

Another exotic type of collective motion is the ``toroidal dipole
mode''. This isoscalar mode, which is characterized by a velocity
field of toroidal shape, was predicted many years ago
\cite{Semenko}. Since the restoring force for this kind of motion is
generated by the distortion of the Fermi surface, this mode cannot be
described within hydrodynamic models. Studies of this mode were
carried out within the method of Wigner function moments
\cite{Balbutsev}, within nuclear fluid dynamics \cite{Bastrukov1993},
and within the relativistic RPA \cite{VretenarPRC}. Since the toroidal
dipole mode is isoscalar and exists also in $N=Z$ nuclei, it was
usually not connected with the pygmy resonance, although different
calculations \cite{VretenarNPA,Ryezayeva} showed that the velocity
field of the pygmy mode has a toroidal shape.

 In the past, semiclassical approaches such as the
Steinwedel-Jensen model \cite{SteinwedelJensen} contributed a lot to
the understanding of giant resonances and how they can be related to
global properties of nuclei. Since semiclassical approaches average
over shell effects \cite{RingSchuck}, they lead to clear and intuitive
pictures which, contrary to the results of quantum mechanical RPA
calculations, are not obscured by details of the specific
single-particle energies and wave functions of the nuclei under
consideration. This is why new modes of excitation, such as the torus
mode mentioned before, were often first identified in semiclassical
approaches.

In the case of the pygmy mode, a generally accepted picture is still
missing. It is not even clear whether the pygmy mode is a generic
collective mode like the giant resonances, or whether it depends on a
particular structure of the single-particle levels. The aim of the
present paper is therefore to study this mode by solving the
semiclassical Vlasov equation for the case of neutron-rich nuclei. The
Vlasov equation has been shown to give a reasonable description of
generic properties of different collective modes of nuclei
\cite{Brink,Burgio,BertschDasGupta}. As we will see, the main result
of the present study is that the pygmy mode is closely related to a
low-lying isoscalar torus mode, which is clearly collective and whose
existence is not limited to nuclei with neutron excess.

In this paper, we are looking for a numerical solution of the full
Vlasov equation without any additional simplifying assumptions, as
opposed to, e.g., fluid-dynamical approaches. A solution of the Vlasov
equation for collective modes was given in \Refs{Brink,Burgio}, but
the calculations were restricted to Woods-Saxon potentials with
separable residual interactions. This is not sufficient for the
description of modes which possibly depend on exotic ground state
properties such as the neutron skin. In the present work, the starting
point is the ground state in Thomas-Fermi (TF) approximation,
calculated self-consistently with the same interaction that is also
used in the Vlasov dynamics. The importance of a consistent
description of the ground state and of the dynamics within a transport
model was recently pointed out in \Ref{Gaitanos} in the context of the
monopole mode.

For the calculation of the mean field entering both the TF and the
Vlasov calculations, an effective interaction capable of describing
exotic nuclei is needed. In the present work, a simplified version of
the so-called BCP functional \cite{Baldo} will be used, which is an
energy functional whose bulk part is based on a fit to microscopic
Br\"uckner calculations, reproducing the equation of state of nuclear
matter in a range of asymmetries from symmetric matter to pure neutron
matter and in a range of densities from zero to more than saturation
density. However, it seems unlikely that the general findings
presented here depend on the details of the interaction.

The paper is organized as follows. In \Sec{sec:method}, the method is
explained. In \Sec{sec:pygmy}, results for the electric dipole
response are presented. Besides the strength function, the transition
densities and velocity fields of the GDR and the PDR are discussed. In
\Sec{sec:torus}, results for the response to the isoscalar toroidal
dipole operator are shown. In particular, by comparing the transition
densities and velocity fields with those discussed in \Sec{sec:pygmy},
we see that in neutron-rich nuclei, the PDR and the GDR are modes
which are excited by both the electric dipole and the isoscalar
toroidal dipole operators due to the mixing of isoscalar and isovector
modes. \Sec{sec:summary} is devoted to the summary and conclusions.

\section{Method}
\label{sec:method}
\subsection{Vlasov equation}
Like other collective vibrations, the pygmy resonance has been studied
within the random-phase approximation (RPA)
\cite{Catara,VretenarNPA,Pena}, which can be interpreted as the
small-amplitude limit of the time-dependent Hartree-Fock (TDHF) theory
\cite{RingSchuck}. Written in terms of the one-body density matrix
$\hat{\rho}$ and the mean-field hamiltonian $\hat{h}$, the TDHF
equation reads\footnote{Protons and neutrons have of course different
  density matrices $\hat{\rho}_p$ and $\hat{\rho}_n$, different
  mean-field hamiltonians $\hat{h}_p$ and $\hat{h}_n$, etc. In order
  to improve the readability, isospin indices $\alpha = p,n$ are
  omitted in this paper except when they cannot be avoided.}
\begin{equation}
i\hbar \dot{\hat{\rho}} = \left[\hat{h},\hat{\rho}\right]\,.
\label{TDHF}
\end{equation}
In the semiclassical $\hbar\to 0$ limit, the TDHF equation reduces to
the Vlasov equation \cite{RingSchuck,BertschDasGupta}. In order to see
this, it is useful to work with the Wigner transforms of $\hat{\rho}$
and $\hat{h}$, which are the distribution function $f(\rv,\pv,t)$ and
the classical mean-field hamiltonian $h(\rv,\pv,t)$. For example, in
the case of a purely local mean field $U$, the latter can be written
as
\begin{equation}
h(\rv,\pv,t) = \frac{p^2}{2m}+U(\rv,t)\,.
\end{equation}
Note that the mean field $U$ depends on time through the time
dependence of the density. To leading order in $\hbar$, the Wigner
transform of the commutator in \Eq{TDHF} reduces to the Poisson
bracket of the corresponding Wigner transforms, and one obtains
the Vlasov equation
\begin{equation}
\dot f = \{h,f\} =
  \frac{\partial h}{\partial \rv}\cdot \frac{\partial f}{\partial \pv}-
  \frac{\partial h}{\partial \pv}\cdot \frac{\partial f}{\partial \rv}\,.
\label{Vlasov}
\end{equation}
A discussion of the Vlasov equation as limiting case of more general
transport equations can be found in \Ref{BotermansMalfliet}.
\subsection{Numerical method}
\label{subsec:numericalmethod}
In order to solve the Vlasov equation (\ref{Vlasov}) numerically, we
will employ the test-particle method which has often been used for the
description of heavy-ion collisions
\cite{BertschDasGupta,Gregoire}. The basic idea of this method is to
replace the distribution function $f(\rv,\pv,t)$ by a finite number of
delta functions (``test particles'')
\begin{equation}
f(\rv,\pv,t) = \frac{1}{\calN} \sum_{i=1}^{\calN A}
\delta(\rv-\rv_i(t))\delta(\pv-\pv_i(t))\,,
\label{ftestparticle}
\end{equation}
where $\calN$ denotes the number of test particles per nucleon and $A$
is the mass number of the nucleus. In order to satisfy the Pauli
principle, the density of test particles of each species ($\alpha =
n,p$) in phase-space must not exceed $2\calN/(2\pi\hbar)^3$ (the
factor of 2 is the spin degeneracy). Inserting \Eq{ftestparticle} into
\Eq{Vlasov}, one finds that each test particle has to follow its
classical trajectory given by
\begin{equation}
\dot{\rv}_i = \frac{\partial h(\rv_i,\pv_i,t)}{\partial \pv_i}\,,\quad
\dot{\pv}_i = -\frac{\partial h(\rv_i,\pv_i,t)}{\partial \rv_i}\,.
\label{traj_general}
\end{equation}
In the case of a purely local mean field, the equations of motion
reduce to
\begin{equation}
\dot{\rv}_i = \frac{\pv_i}{m}\,,
\quad
\dot{\pv}_i = -\nablav U(\rv_i,t)\,.
\label{traj_simple}
\end{equation}
It is clear that these equations of motion prevent the test particles
from entering the classically fobidden region. Note that this absence
of tunneling is inherent to the Vlasov equation and independent of the
numerical method.

The density $\rho(\rv,t)$ corresponding to the distribution function
(\ref{ftestparticle}) is a sum of delta functions and hence not
suitable for any practical calculation. In order to obtain a
well-defined density which can be used, e.g., for the calculation of
the mean field $U$, it is common to replace the delta functions in
\Eq{ftestparticle} by Gaussians \cite{Gregoire,Fennel}, leading to a
smooth density
\begin{equation}
\tilde{\rho}(\rv,t) 
  = \sum_{i=1}^{\calN A} 
    \frac{e^{-(\rv-\rv_i(t))^2/d^2}}{\calN (\sqrt{\pi}d)^3}\,.
\label{gaussian}
\end{equation}

For the sake of consistency, if one uses the smooth density
$\tilde{\rho}(\rv)$ in the calculation of the mean field $U(\rv)$, one
has to modify also the acceleration equation and replace the force at
$\rv_i$ by a force averaged over the Gaussian \cite{Gregoire,Fennel},
\begin{equation}
\dot{\pv}_i = -\nablav\tilde{U}(\rv_i,t)\,,
\end{equation}
where
\begin{equation}
\tilde{U}(\rv,t) = \int \frac{d^3 s}{(\sqrt{\pi}d)^3} e^{-s^2/d^2} 
  U(\rv-\sv,t)\,.
\label{foldU}
\end{equation}

Contrary to quantum molecular dynamics (QMD) \cite{Aichelin} and
related approaches, where similar Gaussians (in $\rv$ and $\pv$ space)
are used to simulate quantum effects, we wish to stay here in the
semiclassical framework and therefore do not attach any profound
meaning to the smoothed density $\tilde{\rho}$. We consider it as an
auxiliary quantity one has to introduce in the test-particle approach
in order to be able to calculate well-defined densities and mean
fields. As we will see in the next subsection, an interesting aspect
of the smoothing of densities and mean fields is that it acts exactly
like a finite-range interaction.

\subsection{Interaction}
\label{subsec:interaction}
Until now, the mean field $U$ entering the hamiltonian $h$ has not
been specified. On the one hand, it should be local for simplicity,
and on the other hand, it should not be too simplistic if one wants to
describe exotic nuclei. In this work, it will be derived from the bulk
part $E_\mathrm{int}^\infty[\rho_p,\rho_n]$ of the
Barcelona-Catania-Paris (BCP) energy functional \cite{Baldo}, which is
a parametrization of Br\"uckner G-matrix results for nuclear and
neutron matter. If the interaction energy is written as
\begin{equation}
E_\mathrm{int}^\infty[\rho_p,\rho_n] = \int d^3 r\,
\epsilon_\mathrm{int}^\infty(\rho_p,\rho_n)\,,
\end{equation}
the mean fields for protons ($\alpha = p$) and neutrons ($\alpha = n$)
are given by
\begin{equation}
U_\alpha(\rho_p,\rho_n) = \frac{\partial \epsilon_\mathrm{int}^\infty}{\partial
  \rho_\alpha}\,.
\end{equation}
As explained in \Sec{subsec:numericalmethod}, $U$ is calculated from
the smoothed densities $\tilde{\rho}$, i.e.,
\begin{equation}
U_\alpha(\rv,t) = U_\alpha(\tilde{\rho}_p(\rv,t),\tilde{\rho}_n(\rv,t))\,.
\label{meanU}
\end{equation}

In addition to the bulk part $E_\mathrm{int}^\infty$, the BCP
functional contains a finite-range part $E_\mathrm{int}^\mathrm{FR}$,
a spin-orbit part $E^{\mathrm{s.o.}}$, and a Coulomb part
$E_{\mathrm{C}}$. The finite-range part, which was introduced in
\Ref{Baldo} in order to get the right surface energy, has mainly the
effect to smooth out the mean fields as compared to the densities.
The same effect can be achieved without any additional term
$E_\mathrm{int}^\mathrm{FR}$ if one introduces a finite range into the
bulk term $E_\mathrm{int}^\infty$. For instance, in the latest version
of the BCP functional \cite{Robledo}, $E_\mathrm{int}^\mathrm{FR}$ has
been substituted by a finite range in the quadratic term of
$E_\mathrm{int}^\infty$.

Here, instead of implementing directly a finite-range force between
the test particles, we first calculate the smooth density by folding
the distribution function with a Gaussian of width $d$
[\Eq{gaussian}]. From this smoothed density, we calculate the mean
field [\Eq{meanU}] which is folded again by a Gaussian of width $d$
when the force on a particle is calculated [\Eq{foldU}]. If
$\epsilon_\mathrm{int}^\infty$ was quadratic in the densities, this
procedure would be equivalent to a Gaussian finite-range interaction
with width $\sqrt{2} d$. Hence, if we choose $d = 0.7$ fm, this
corresponds roughly to the range $r_0 = 1.05$ fm \cite{Baldo} of the
finite-range term in the BCP functional.

For the sake of simplicity, the spin-orbit and Coulomb contributions,
$E^{\mathrm{s.o.}}$ and $E_{\mathrm{C}}$, will be neglected in the
present work. Note that, according to the Kohn-Sham energy density
functional theory, the BCP energy functional does not introduce an
effective mass $m^*$ in the kinetic energy part. Hence, the mass $m$
which appears in the hamiltonian $h$ is the free nucleon mass.

With the present prescription to calculate the force on a test
particle, it is straight-forward to show that the total energy defined
by
\begin{equation}
E_\mathrm{tot} = \frac{1}{\calN} \sum_{i=1}^{\calN A} \frac{p_i^2}{2m}
  + E_\mathrm{int}^\infty[\tilde{\rho}_p,\tilde{\rho}_n]
\end{equation}
is exactly conserved during the time evolution.
\subsection{Ground state initialization}
\label{subsec:initialization}
For a given initial distribution function $f(\rv,\pv,t=t_0)$, the time
evolution at $t > t_0$ is completely determined by \Eq{Vlasov}. Here,
we choose as initial state an isolated nucleus at rest in its ground
state. An obvious requirement for the ground state is that it must be
stationary. In the present framework, this means that the ground state
must be calculated within the TF approximation, $f(\rv,\pv) =
\theta(\mu-h(\rv,\pv))$, which is stationary under the Vlasov equation
(\ref{Vlasov}). Here, $\theta$ is the step function and $\mu$ is the
chemical potential (Fermi energy). For consistency, since the
numerical simulation of the Vlasov equation requires to smooth the
densities and mean fields with Gaussians, \Eqs{gaussian} and
(\ref{foldU}), the same Gaussians should be included in the
calculation of the TF ground state \cite{Fennel}.

Written explicitly, the equations which have to be solved
self-consistently are 
\begin{gather}
\rho(\rv) =
\frac{[2m(\mu-\tilde{U}(\rv))]^{3/2}}{3\pi^2\hbar^3}
  \theta(\mu-\tilde{U}(\rv))\,, \label{rhoTF}\\
\tilde{\rho}(\rv) = \int \frac{d^3 s}{(\sqrt{\pi}d)^3} e^{-s^2/d^2} 
  \rho(\rv-\sv)\,,\label{foldrho}
\end{gather}
where $\tilde{U}$ is computed from $\tilde{\rho}$ according to
\Eqs{meanU} and (\ref{foldU}). The chemical potentials (Fermi
energies) $\mu_p$ and $\mu_n$ are determined from the conditions
\begin{equation}
Z = \int d^3r \rho_p(\rv)\,,\quad N = \int d^3r \rho_n(\rv)\,.
\end{equation}

\begin{figure}
\centering{\includegraphics[height=3.9cm]{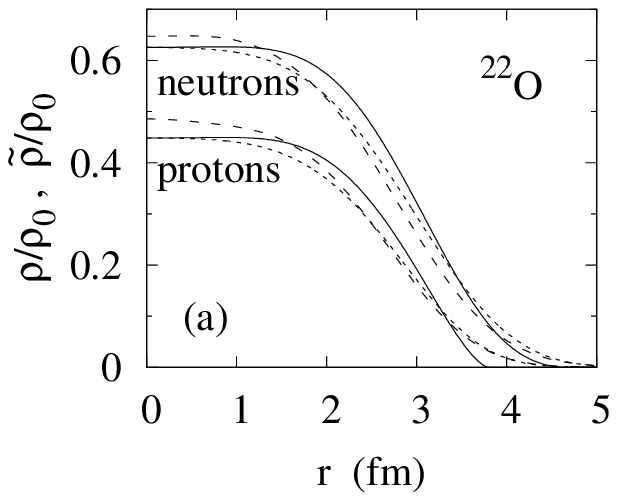}
\includegraphics[height=3.9cm]{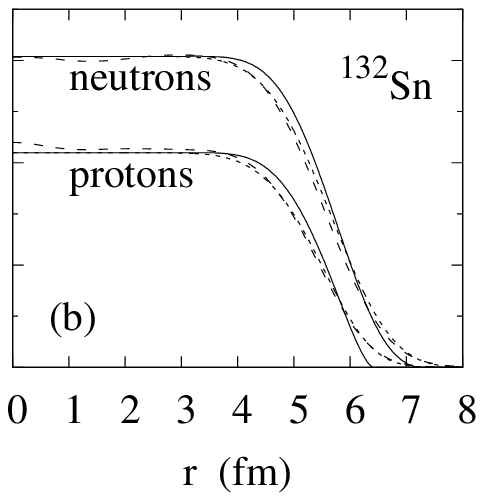}}
\caption{{Ground-state densies of neutrons and protons in units of
    $\rho_0=0.17$ fm$^{-3}$ in $^{22}$O (a) and $^{132}$Sn (b). Solid
    lines: self-consistent TF densities $\rho(r)$, dots: corresponding
    smoothed densities $\tilde{\rho}(r)$ according to \Eq{foldrho},
    dashes: smoothed densities $\tilde{\rho}(r)$ after a simulation
    time of $t = 2000$ fm$/c$.}}
\label{figure1}
\end{figure}

For illustration, the density distributions of protons and neutrons in
$^{22}$O and $^{132}$Sn obtained in this way are displayed in
\Fig{figure1}.
The neutron skin is clearly visible in both cases.  Note that the
TF densities (solid lines) vanish at the classical turning points
which are determined by $\mu_\alpha = \tilde{U}_\alpha$. The finite
surface thickness stems from the self-consistent solution of the TF
equations as described by \Eqs{rhoTF}, (\ref{foldrho}), (\ref{meanU}),
and (\ref{foldU}) with $d=0.7$ fm.

Once the self-consistent TF density distributions are obtained, the
test particle positions $\rv_i$ are initialized randomly according to
a probability density $P(\rv_i) \propto \rho(\rv_i)$. Then, the
momenta $\pv_i$ are initialized randomly in a sphere with radius
$p_F(\rv_i) = \hbar(3\pi^2\rho(\rv_i))^{1/3}$ in order to correctly
describe the Fermi motion. Remember that, if the Pauli principle is
satisfied initially, it is preserved by the Vlasov dynamics due to
Liouville's theorem \cite{Gregoire}.
\subsection{Numerical parameters and stability}
As mentioned before, the width $d$ of the Gaussians has two
effects: first, it is necessary to obtain a well-defined density
distribution, and second, it induces effectively a finite-range in the
interaction. Most of the results to be presented in this paper
were obtained with $d = 0.7$ fm, leading to a reasonable smoothing of
the mean field as discussed in \Secs{subsec:interaction} and
\ref{subsec:initialization}. This defines the minimum number of test
particles to be used, as there must be a sufficiently large number of
test particles per volume $d^3$, otherwise the statistical
fluctuations become too strong. Here, $\calN = 2000$ test particles
per particle were used. The equations of motion were solved with the
velocity Verlet algorithm \cite{Swope} using a time step of 0.1
fm$/c$. After each time step, the mean field $U$ was updated and
stored in a three dimensional grid with spacing 0.4 fm.

With these parameters, it was possible to ensure the stability of
nuclei with large neutron excess, i.e., with very weakly bound
neutrons in the surface, during the long simulation time of 2000
fm/$c$ which is necessary for the calculation of the response function
(see next subsection). The numerical losses due to test particles
which escape from the nucleus are $< 1.3$ neutrons in the case of
$^{22}$O and $< 4$ neutrons in the case of $^{132}$Sn. In order to
illustrate the stationarity of the ground state, we display in
\Fig{figure1} the angle-averaged (see appendix) smoothed densities
$\tilde{\rho}$ corresponding to the test-particle distribution after
the simulation (dashes), which agree very well with those calculated
during the initialization (dots). During the simulation, the kinetic
energy (including that carried away by the test particles escaping
from the nucleus) drops by $\sim$ 5\% in the case of $^{22}$O and by
$\sim$ 2\% in the case of $^{132}$Sn. The relative variation of the
total (kinetic plus interaction) energy is of the order of $10^{-6}$.

In order to check that the results do not depend sensitively on the
value of the width parameter $d$, some calculations with $d=0.5$ fm
were performed. In this case, the number of test particles per
particle was increased to $\calN = 5000$ in order to limit statistical
fluctuations, and the spacing of the grid for the mean field was
reduced to 0.3 fm.
\subsection{Calculation of the response function}
\label{subsec:responsefunction}
Within the present approach, the collective modes are described as
time-dependent oscillations of the nucleus after a perturbation of the
ground state. In order to relate these oscillations to the usual
response function, which is defined in terms of transition
probabilities from the ground state to excited states, let us
temporarily leave the semiclassical framework and return to quantum
mechanics. We consider a perturbation hamiltonian of the form
$\hat{H}_\mathrm{ex}(t) = \lambda \hat{Q} \delta(t)$, where $\hat{Q}$
is the excitation operator we want to study and $\lambda$ is supposed
to be small. Then, within linear response theory \cite{FetterWalecka},
the expectation value of the operator $\hat{Q}$ as a function of time
is given by
\begin{align}
\delta\langle \hat{Q}\rangle(t) &= \langle \hat{Q}\rangle(t)-\langle
  0|\hat{Q}|0\rangle \nonumber \\ &= -\frac{2 \lambda\theta(t)}{\hbar}
  \sum_f |\langle f|\hat{Q}|0\rangle|^2 \sin\frac{(E_f-E_0)t}{\hbar}\,,
\end{align}
where $|0\rangle$ is the ground state (of the unperturbed hamiltonian
$\hat{H}$), $|f\rangle$ is an excited state, and $E_0$ and $E_f$ are
the corresponding energies. Defining the strength function as usual by
\begin{equation}
S(E) = \sum_f |\langle f|\hat{Q}|0\rangle|^2 \delta(E-E_f+E_0)\,,
\end{equation}
we can obtain it from $\delta\langle \hat{Q}\rangle(t)$ via a Fourier
transform
\begin{equation}
S(E) = -\frac{1}{\pi\lambda}\int_0^\infty dt
  \delta\langle\hat{Q}\rangle(t) \sin\frac{E t}{\hbar}\,.
\label{responsefourier}
\end{equation}

Let us now return to the semiclassical framework. Under the assumption
that $\hat{Q}$ is a one-body operator, i.e.
\begin{equation}
\hat{Q} = \sum_{i=1}^A \hat{q}_i\,.
\end{equation}
one can calculate its expectation value as
\begin{equation}
\langle \hat{Q}\rangle(t) = \int d^3 r\, d^3 p\, f(\rv,\pv,t) q(\rv,\pv)\,,
\end{equation}
where $q(\rv,\pv)$ is the Wigner transform of $\hat{q}$, which can be
obtained (at least to leading order in the $\hbar$ expansion
\cite{RingSchuck}) by replacing the operators $\hat{\rv}$ and
$\hat{\pv}$ in $\hat{q}$ by their classical counterparts $\rv$ and
$\pv$. In terms of the test particle positions and momenta, this
expectation value can be expressed as
\begin{equation}
\langle \hat{Q}\rangle(t) = \frac{1}{\calN}\sum_{i=1}^{\calN A}
   q(\rv_i(t),\pv_i(t))\,.
\end{equation}

The last point which remains to be explained is how the delta function
perturbation at $t = 0$ changes the initial distribution function,
i.e., in our simulation, the distribution of test particles. In
principle, one has to solve the classical equations of motion
(\ref{traj_general}) with the perturbed hamiltonian $h + \lambda q
\delta(t)$ instead of $h$. Replacing the delta function by a short
pulse of length $\delta t$, one can show that in the limit $\delta
t\to 0$ and to leading order in $\lambda$, the effect of the
perturbation is to change the positions and momenta of the test
particles as follows\footnote{If $q$ depends only on $\rv$ or only on
$\pv$, this result is valid to all orders in $\lambda$.}:
\begin{equation}
\rv_i \to \rv_i +\lambda \frac{\partial q(\rv_i,\pv_i)}
  {\partial \pv_i}\,, \quad 
\pv_i \to \pv_i -\lambda \frac{\partial  q(\rv_i,\pv_i)}
  {\partial \rv_i}\,.
\label{kick}
\end{equation}

To summarize the procedure: First, the test-particle distribution at
$t = 0$ is initialized as explained in \Sec{subsec:initialization}. Then
the test-particle positions and momenta are changed according to
\Eq{kick} and the mean field $U$ is recalculated if necessary (if the
excitation operator $q$ depends on $\pv$). After that, the equations
of motion (\ref{traj_simple}) are solved simultaneously for all test
particles, and the mean field $U$ is updated after each time step. In
this way, one obtains the expectation value $\langle
\hat{Q}\rangle(t)$ as a function of time, and its Fourier transform
(\ref{responsefourier}) gives the strength function $S(E)$.

In practice, it is of course impossible to run the simulation to $t =
\infty$. Here, the simulations will be stopped at $t_\mathrm{max} =
2000$ fm$/c$. In order to avoid oscillations in the Fourier transform
related to the cut at $t_\mathrm{max}$, the strength function $S(E)$
will be folded with a Lorentzian of width $\gamma = 0.5$ MeV, which is
equivalent to multiplying the sine function in \Eq{responsefourier} by
$e^{-\gamma t/2\hbar}$.

\subsection{Transition densities and velocity fields}
The delta function perturbation at $t = 0$ excites simultaneously all
modes which can be excited by the operator $\hat{Q}$. It is therefore
difficult to extract the transition density and velocity field
corresponding to one particular mode. What can be done is to calculate
the (smoothed) density distributions ${\tilde{\rho}}(\rv,t)$ and
velocities
\begin{equation}
\vv(\rv,t) =
\frac{{\tilde{\jv}}(\rv,t)}{\tilde{\rho}(\rv,t)} =
\frac{1}{\tilde{\rho}(\rv,t)}
\sum_{i=1}^{\calN A} \frac{\pv_i(t)}{m}\,
\frac{e^{(\rv-\rv_i(t))^2/d^2}}{\calN (\sqrt{\pi}d)^3}
\end{equation}
(see appendix for more details) as functions of time.

In the case of a time-even excitation operator, i.e., $q(\rv,\pv) =
q(\rv,-\pv)$, the particles get a kick in momentum space and the
oscillation starts with maximum velocity, while the density is not
changed at $t = 0$. The situation is opposite if the excitation
operator is time-odd, i.e., $q(\rv,\pv)=-q(\rv,-\pv)$: In this case,
the velocity is zero at $t = 0$, while the density is immediately
changed due to the displacement of the particles in coordinate space.

In order to find the contribution of a given mode to the density and
velocity oscillations, one has to choose the energy $E$ corresponding
to a peak in the strength function and compute the transition
densities and velocity fields as a Fourier transform of
$\tilde{\rho}(\rv,t)$ and $\vv(\rv,t)$, respectively. In the case of a
time-even excitation operator, one has to use
\begin{gather} \delta \tilde{\rho}(\rv,E) \propto
  \int_0^\infty dt  \tilde{\rho}(\rv,t) 
    \sin\frac{E t}{\hbar}\,,\\
\vv(\rv,E) \propto \int_0^\infty dt \vv(\rv,t)
    \cos\frac{E t}{\hbar}\,,
\end{gather}
whereas in the case of a time-odd operator, the sine and cosine
functions must be interchanged. In practice, since the simulation runs
only to $t_\mathrm{max} = 2000$ fm$/c$, the sine and cosine
functions are multiplied by an exponential damping factor $e^{-\gamma
t/2\hbar}$, $\gamma = 0.5$ MeV, as in the strength function $S(E)$.

It should be noted that even if the energy $E$ corresponds to a peak
in $S(E)$, the transition densities and velocity fields obtained with
this method may still contain contributions from other modes if those
have a width which makes their spectrum extend to energy $E$.

\section{Results for the pygmy resonance}
\label{sec:pygmy}
\subsection{Electric dipole response}
\label{subsec:electric}
Since the pygmy resonance is often studied in $(\gamma,\gamma^\prime)$
experiments, we consider as excitation operator the electric dipole
operator \cite{BohrMottelson}
\begin{equation}
q =
  \begin{cases} \frac{N}{A} z & \mbox{for protons},\\ -\frac{Z}{A} z
  & \mbox{for neutrons},
\end{cases}
\label{eldipole}
\end{equation}
which is defined such that the center of mass of the nucleus stays at
rest. For the parameter $\lambda$ multiplying the operator $q$, the
value $\lambda = 25$ MeV$/c$ is chosen as a compromise to excite an
oscillation which is much larger than the numerical noise due to the
finite number of test particles, but still small enough so that
nonlinearities do not play a role. Calculations for different nuclei
from oxygen isotopes up to $^{208}$Pb were performed. The existence of
the PDR as a small enhancement in the strength function well below the
energy of the GDR turned out to be a general property of $N > Z$
nuclei, while it is absent in $N=Z$ nuclei.

\begin{figure}
\centering{\includegraphics[width=8cm]{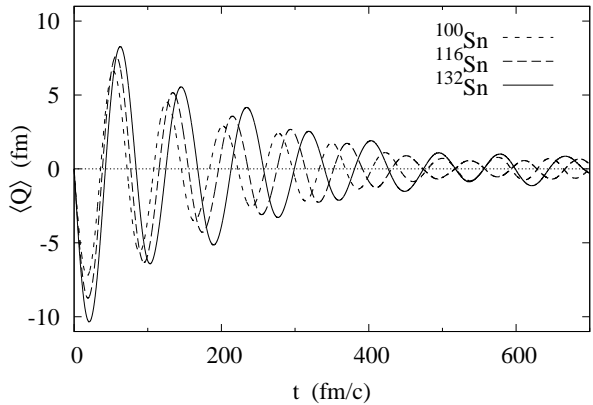}}
\caption{Electric dipole moment of $^{100}$Sn, $^{116}$Sn, and
$^{132}$Sn after a perturbation with $\hat{H}_\mathrm{ex}(t) = \lambda
\hat{Q} \delta(t)$, $\hat{Q}$ being the electric dipole operator and
$\lambda = 25$ MeV$/c$.}
\label{figure2}
\end{figure}

\begin{figure}
\centering{\includegraphics[width=8cm]{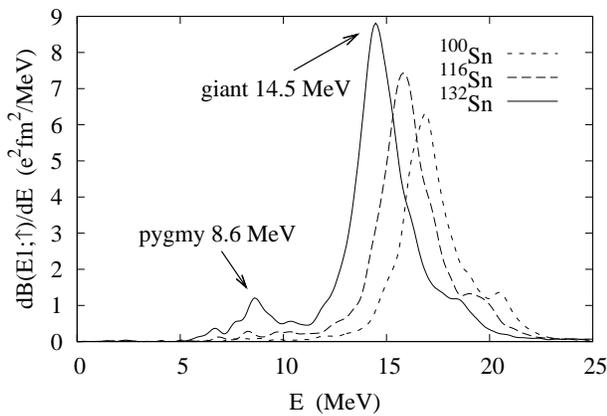}}
\caption{Electric dipole strength of $^{100}$Sn, $^{116}$Sn, and
$^{132}$Sn.}
\label{figure3}
\end{figure}

As a first example, let us discuss the results obtained for the three
tin isotopes $^{100}$Sn, $^{116}$Sn, and $^{132}$Sn. In all three
cases, after initializing and exciting the nucleus, one observes a
damped oscillation of $\langle \hat{Q}\rangle (t)$ with the frequency
of the GDR, see \Fig{figure2}.
The curves look qualitatively similar for all three nuclei, and in
order to see anything else than the GDR, one has to look at their
Fourier transforms. In \Fig{figure3}, the corresponding electric
dipole strengths are displayed.
In all three nuclei, there is a very strong peak at around 14.5 MeV
($^{132}$Sn) to 16.7 MeV ($^{100}$Sn) corresponding to the GDR. The
energies of the GDR are in reasonable agreement with experimental
values, whereas the widths are much too small. For instance, the curve
shown for $^{116}$Sn is very well fitted by a Lorentzian with energy
15.8 MeV and width 2.4 MeV (which includes the artificial width
$\gamma = 0.5$ MeV mentioned in the end of
\Sec{subsec:responsefunction}), whereas the corresponding experimental
energy and width are 15.67 and 4.19 MeV, respectively \cite{Fultz}.

It is in fact not surprising that the widths are too small. Since in
the linear regime the Vlasov equation is a semiclassical version of
the RPA \cite{Brink,Burgio}, the only damping mechanism that is
present here is Landau damping. The fact that in an anharmonic
potential different classical orbits have different periodicities
leads to effects which are completely analogous to the splitting of
the quantum mechanical single-particle levels, as already noticed in
\Refs{Brink,Burgio}. The main difference to quantum mechanical RPA
calculations is that within Vlasov dynamics the strength is not
fragmented into many discrete states, but it is
continuous\footnote{Note that in \Refs{Brink,Burgio} the angular
  momentum of the classical orbits was artificially quantized in order
  to simplify the practical calculations and to obtain a discrete
  spectrum as in RPA. In the present work, the angular momentum of the
  test particles is arbitrary, which results in a continuous
  spectrum.}. The fragmentation due to the coupling to more complex
states like two-phonon or two-particle-two-hole states is missing
here, as it is in RPA \cite{BertschBortignon}. In the semiclassical
framework, effects analogous to two-particle-two-hole excitations can
be included via a collision term \cite{Burgio}, but this is beyond the
scope of the present work.

Let us return to the discussion of the results shown in
\Fig{figure3}. In the case of $^{116}$Sn, one can see a small amount
of dipole strength below the GDR which is absent in the $N=Z$ nucleus
$^{100}$Sn and which becomes a well defined peak at 8.6 MeV in the
case of $^{132}$Sn. This peak corresponds to the PDR. For comparison,
in experiment, it was seen in $^{130}$Sn and $^{132}$Sn at a slightly
higher energy of approximately 9.8 MeV \cite{Adrich2005}. Subtracting
the tail of the GDR (assuming that it has the same shape as in
$^{100}$Sn), one finds that in $^{132}$Sn the PDR contributes about
4\% to the energy-weighted sum rule (EWSR), which happens to be in
perfect agreement with the experimental value from
\Ref{Adrich2005}. However, the experimental number for $^{130}$Sn is
larger ($7\%$ of the EWSR) than that for $^{132}$Sn although the
neutron excess is smaller, certainly due to the doubly magic nature of
$^{132}$Sn. In a theory without shell effects, the results for
$^{130}$Sn and $^{132}$Sn are of course almost identical. Nevertheless
this comparison shows that, in spite of its crudeness, the
semiclassical approach is capable of giving the right order of
magnitude for the transition strength.

 In order to see how sensitive these results are to the choice of
the width parameter $d$, the calculation for $^{132}$Sn was repeated
with $d = 0.5$ instead of 0.7 fm. The main effect of this change on
the electric dipole response is that the GDR is slightly shifted from
14.5 to 15.3 MeV. The position of the PDR is almost not affected (the
maximum of the peak is shifted from 8.6 to 8.7 MeV). The height of the
peak corresponding to the PDR is slightly reduced, but subtracting the
tail of the GDR, which is now further apart, one finds again that it
contributes about 4\% of the EWSR. The shape of the transition
densities and velocity fields (see next subsection) of the two modes
are not changed either, except that they go to zero more rapidly at
the surface. The conclusion is that the results do not depend strongly
on the parameter $d$, and from now on we will keep $d = 0.7$ fm which
is the value motivated in \Sec{subsec:interaction}.

\begin{figure}
\centering{\includegraphics[width=8cm]{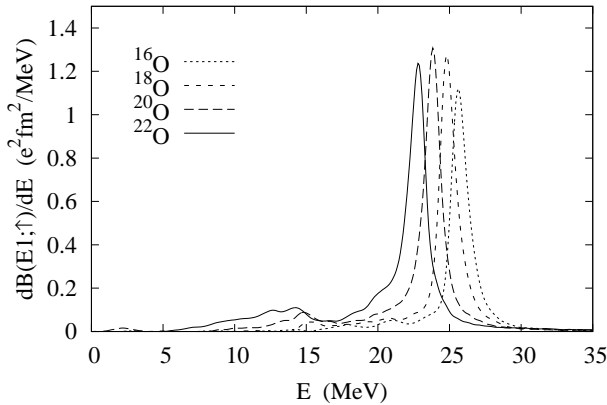}}
\caption{Electric dipole strength of $^{16}$O, $^{18}$O, $^{20}$O, and
  $^{22}$O.}
\label{figure4}
\end{figure}

As a second example let us consider the even oxygen isotopes from
$^{16}$O to $^{22}$O. The corresponding strength functions are
displayed in \Fig{figure4}.
As in the case of tin isotopes, it can be seen that with increasing
neutron excess some strength builds up at low energies, which is
clearly separated from the GDR. Quantitatively, if integrated up to 15
MeV, its contribution to the EWSR is 0\% for $^{18}$O, 4\% for
$^{20}$O, and 8\% for $^{22}$O. This has to be compared with the
corresponding experimental numbers which are 8\% for $^{18}$O, 12\%
for $^{20}$O, and 7\% for $^{22}$O \cite{Leistenschneider}. It is
interesting to notice that, as in the case of $^{130}$Sn and
$^{132}$Sn, the experimental results for the contribution of the
low-lying strength to the EWSR do not increase with increasing neutron
excess. This must be related to shell effects and cannot be reproduced
within the semiclassical framework. Unlike in $^{132}$Sn, the
low-lying strength distribution in $^{22}$O is completely spread and
does not have a clear peak. In this case, it does not seem to be
appropriate to speak of the pygmy resonance as a collective mode.

Finally, let us discuss some numbers for the nucleus $^{208}$Pb. The
response (not shown) looks qualitatively similar to the $^{132}$Sn
case: the pygmy resonance shows up as a well-defined peak. This peak
is situated at 7.6 MeV, while the experimental spectrum has two groups
of transitions around 5.3 and 7.3 MeV \cite{Ryezayeva}. The result for
the total strength $B(E1;\uparrow)$ integrated up to 8 MeV is 2
$e^2$fm$^2$ within Vlasov and 1.32 $e^2$fm$^2$ in the experiment
\cite{Ryezayeva}.

\subsection{Velocity fields and transition densities}
In order to study the nature of the collective modes, it is useful to
look at the transition densities and the velocity fields. The
graphical representation of the transition density can be simplified
by assuming that the amplitude of the oscillation is weak (linear
response regime). In this case, the spherical symmetry of the ground
state and the dipole form of the excitation operator imply that the
transition density can be written as
$\delta \tilde{\rho}(\rv)=\delta{\tilde{\rho}}(r)\cos\theta$,
  where $r = |\rv|$ and $\cos\theta = z/r$.

\begin{figure}
\centering{\includegraphics[width=6cm]{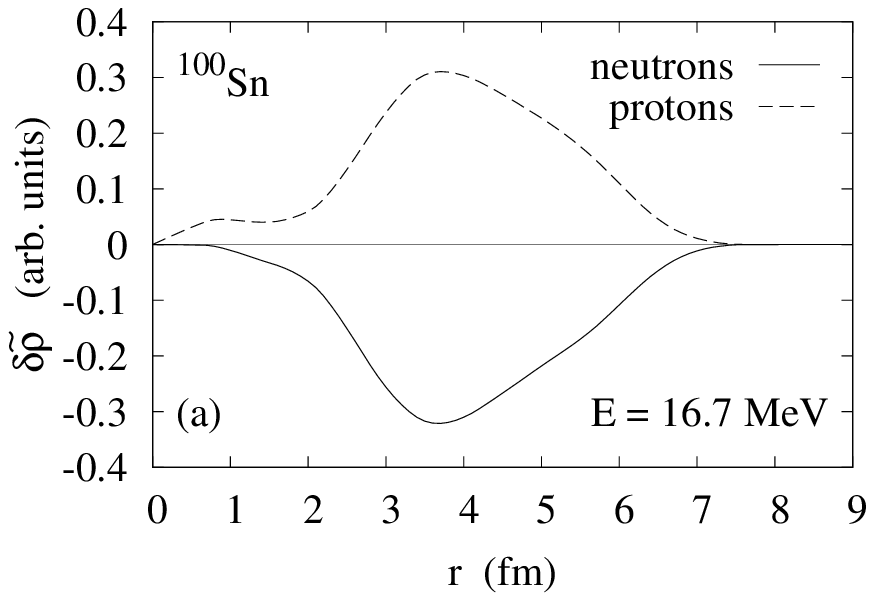}}\\
\centering{\includegraphics[width=6cm]{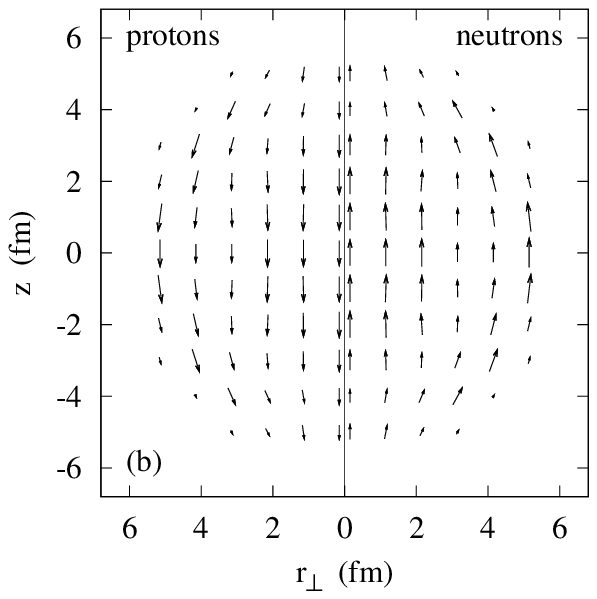}}
\caption{(a) Transition densities $\delta \tilde{\rho}_p$ (dashes) and
  $\delta \tilde{\rho}_n$ (solid line) and (b) velocity fields $\vv_p$
  (left) and $\vv_n$ (right) corresponding to the GDR in $^{100}$Sn.}
\label{figure5}
\end{figure}

In order to test the calculation of transition densities and velocity
fields, let us start with a simple example, namely with the GDR in the
symmetric nucleus $^{100}$Sn. The results are shown in \Fig{figure5}.
As expected, protons and neutrons move against each other in $z$
direction. The velocity (\Fig{figure5}b) is not constant but decreases
with increasing $r$ and gets curved, almost as in the
Steinwedel-Jensen model in which the radial component of the velocity
field vanishes at the surface \cite{SteinwedelJensen}. Since the
Coulomb interaction is not included in the present calculation, the
transition densities and velocity fields of neutrons and protons
should be exactly opposite to each other. For the velocity fields
(\Fig{figure5}b), this seems to be the case, but in the transition
densities (\Fig{figure5}a) a discrepancy is present at small radii
($\lesssim 2$ fm). This is clearly a numerical error. The reason is
that at small radii, the angle averaging, which is implicit in the
computation of the radial functions $\delta \tilde{\rho}(r)$, is
less effective in reducing statistical fluctuations than at large
radii. Since the amplitude of the oscillation is very small, already
small statistical fluctuations of the density can lead to an erroneous
result for the transition density. This is why the transition
densities at $r \lesssim 2$ fm cannot be trusted.

\begin{figure}
\centering{\includegraphics[width=6cm]{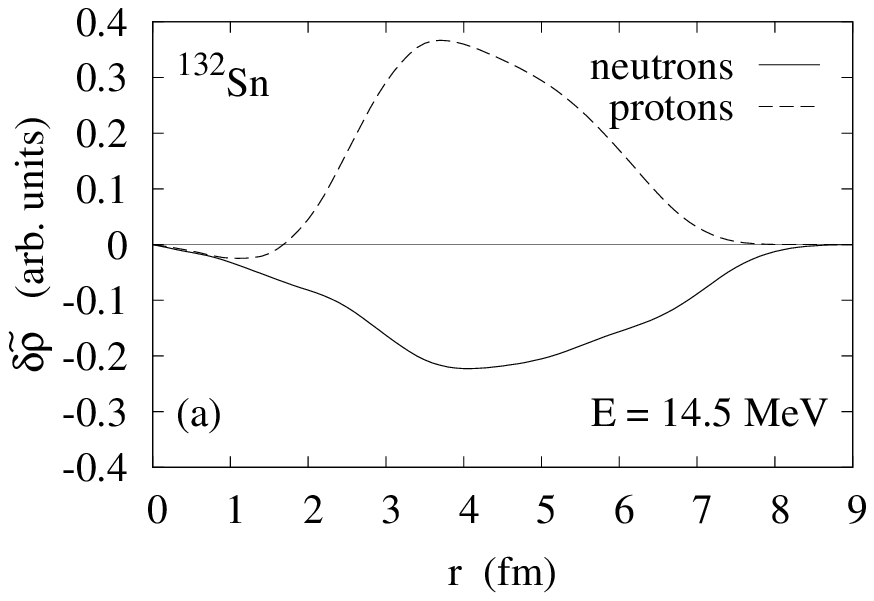}}\\
\centering{\includegraphics[width=6cm]{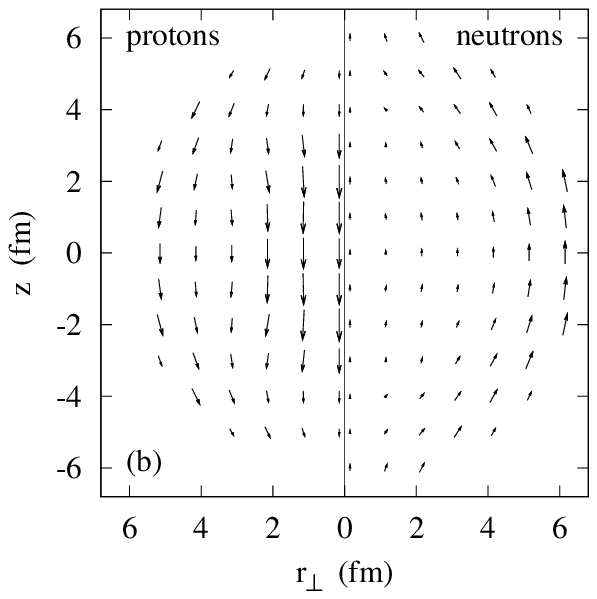}}
\caption{Same as \Fig{figure5}, but for the GDR in $^{132}$Sn.}
\label{figure6}
\end{figure}

After this word of caution, let us look at the more interesting case
of the neutron rich nucleus $^{132}$Sn. Since in this nucleus the
neutron and proton density distributions in the ground state are
different, one does not expect any more that the transition densities
and velocity fields of neutrons and protons are exactly opposite to
each other. Generally speaking, in $N\neq Z$ nuclei, even in the
absence of Coulomb interaction, the collective modes are not exactly
isovector or isoscalar ones, but they have both isovector and
isoscalar components. Let us first discuss the GDR which is displayed
in \Fig{figure6}.
Since the transition densities for $r\lesssim 2$ fm are not reliable,
the node of $\delta \tilde{\rho}_p$ (dashed line in \Fig{figure6}a) is
most likely a numerical error. Beyond that radius, the shape of the
transition densities is typical for the GDR. As a consequence of the
neutron skin, the transition density of neutrons extends to larger
radii than that of protons. The velocity fields (\Fig{figure6}b) are
more surprising: While the proton velocity is very similar to the one
in $^{100}$Sn, the neutron velocity is very different. It seems that
in $^{132}$Sn the neutron velocity is strongly suppressed in the
center and enhanced in the neutron skin. The origin of this phenomenon
is not completely understood, but a possible explanation could be the
coupling between the isovector GDR and the isoscalar torus mode, see
next section.

\begin{figure}
\centering{\includegraphics[width=6cm]{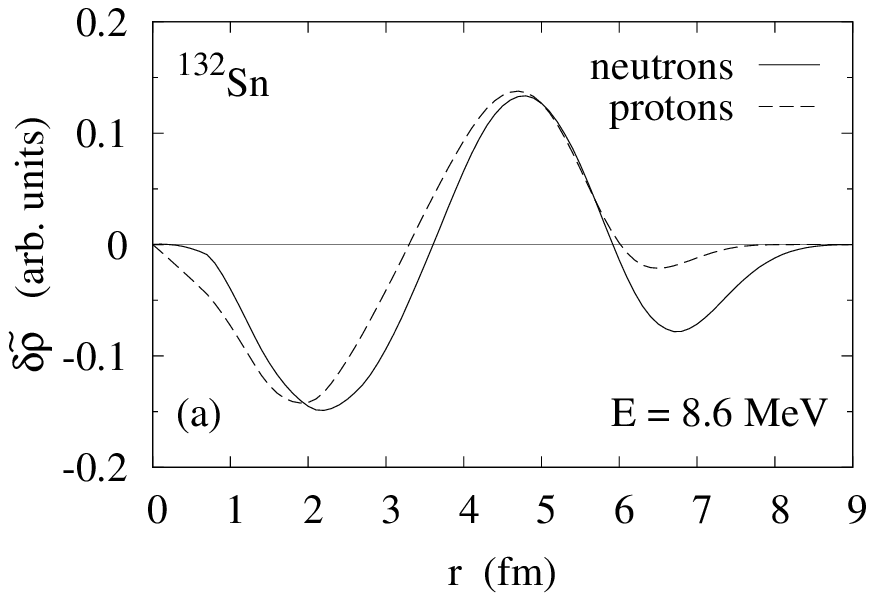}}\\
\centering{\includegraphics[width=6cm]{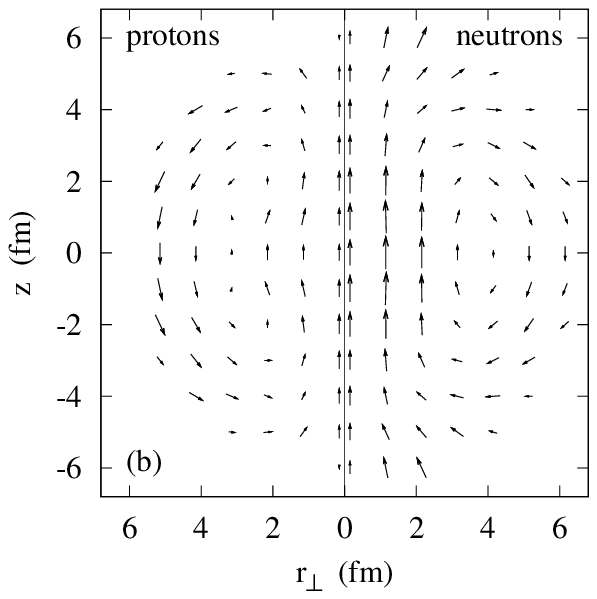}}
\caption{Same as \Fig{figure5}, but for the PDR in $^{132}$Sn.}
\label{figure7}
\end{figure}

In \Fig{figure7},
the transition densities and velocity fields corresponding to the
pygmy mode are displayed. As one can see, protons and neutrons
oscillate mainly in phase. The isovector component of the pygmy mode
comes from the different transition densities in the region of the
neutron skin. The velocity field has a toroidal shape, very different
from the giant resonance. Such a shape was already found in quantum
mechanical calculations of the velocity field of the pygmy mode in
$^{122}$Zr \cite{VretenarNPA} and in $^{208}$Pb \cite{Ryezayeva}. In
the literature, it is often said that the PDR is an oscillation of the
neutron skin against the $N=Z$ core of the nucleus \cite{Paar}. Due to
the toroidal form of the velocity field, the neutrons in the neutron
skin indeed move against the neutrons in the core. However, the image
of the skin oscillating as a whole against an inert core seems to be
over-simplified, since the protons have a toroidal flow-pattern, too.
\section{Results for the torus mode}
\label{sec:torus}
\subsection{Toroidal excitation spectrum}
Motivated by the toroidal shape of the velocity field of the PDR, let
us have a closer look at the isoscalar torus mode. Its excitation
operator is given by
\begin{equation}
q(\rv,\pv) = \left(2r^2-\tfrac{5}{3}\langle r^2\rangle\right)p_z-
  (\pv\cdot\rv)z\,.\label{torusoperator}
\end{equation}
Here, the term $\propto \langle r^2\rangle p_z$ has been added to the
operator given in \Ref{Balbutsev} in order to make sure that the
excitation does not displace the center of mass of the nucleus. Note
that, according to \Eq{kick}, this excitation operator leads to a
change of both positions and momenta of the test particles at $t = 0$
since it depends on $\rv$ and $\pv$.

\begin{figure}
\centering{\includegraphics[width=7cm]{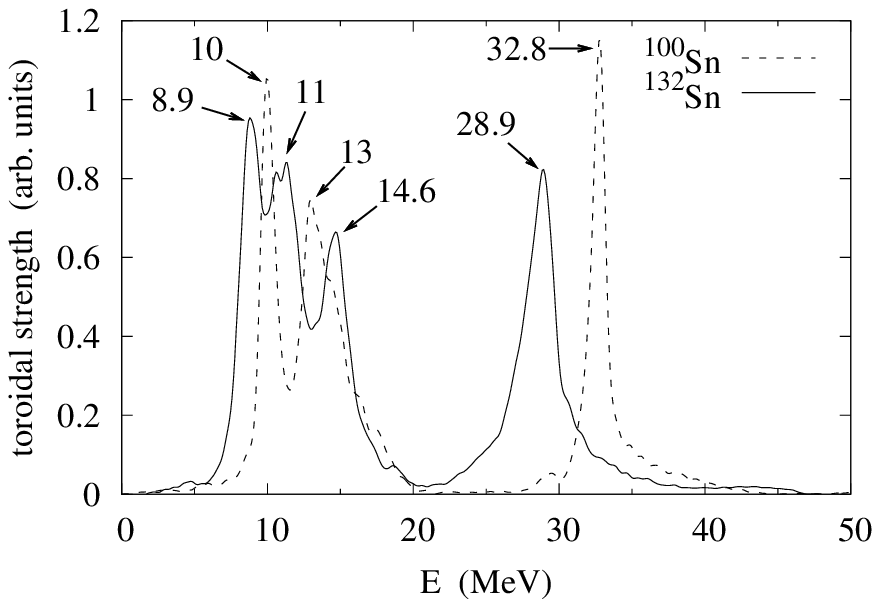}}
\caption{Strength of the response of $^{100}$Sn and $^{132}$Sn to the
  toroidal dipole operator, \Eq{torusoperator}.}
\label{figure8}
\end{figure}

The corresponding strength functions of $^{100}$Sn and $^{132}$Sn are
shown in \Fig{figure8}.
We see that the strength is split into two regions below and above
$\sim 20$ MeV. The first region contains two peaks at $10$ and $13$
MeV in the case of $^{100}$Sn and three peaks at $8.9$, $11$, and
$14.6$ MeV in the case of $^{132}$Sn. In the second region, there is
an isolated peak at $32.8$ MeV in the case of $^{100}$Sn and $28.9$
MeV in the case of $^{132}$Sn. The nature of the different modes will
be clarified by the analysis of the corresponding transition densities
and velocity fields. It is interesting to notice that the positions of
the modes in $^{100}$Sn at $10$ and $32.8$ MeV are in good agreement
with recent RPA results obtained with the UCOM interaction
\cite{Papakonstantinou}. The mode at $13$ MeV corresponds probably to
the fragmented strength concentrated around $15$ MeV in the RPA
response, see Fig. 1 of \Ref{Papakonstantinou}.

\subsection{Velocity fields and transition densities}
\begin{figure}
\centering{\includegraphics[width=6cm]{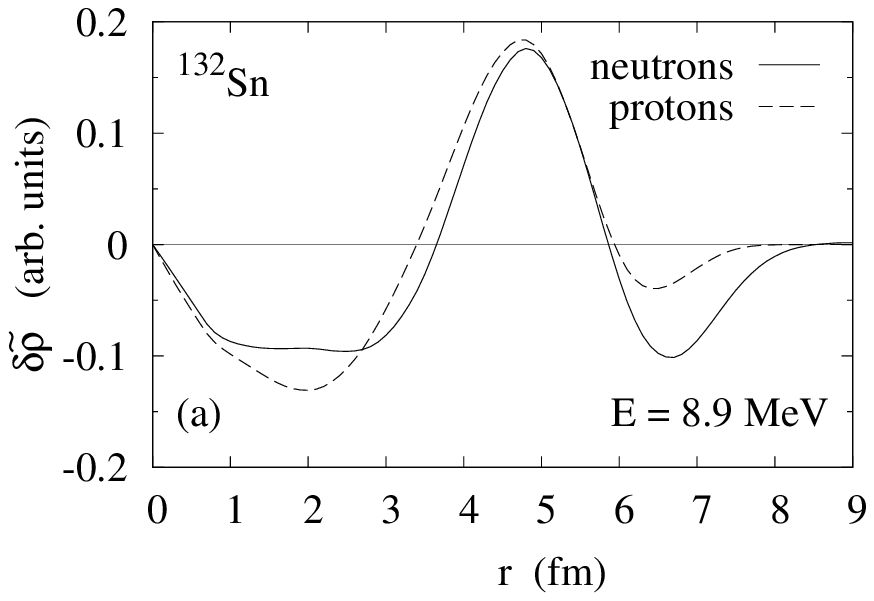}}\\
\centering{\includegraphics[width=6cm]{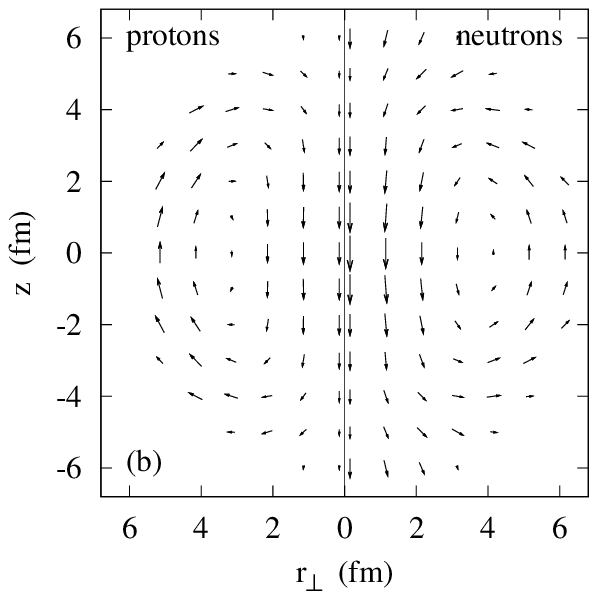}}
\caption{Same as \Fig{figure5}, but for the mode excited by the
toroidal dipole operator at $8.9$ MeV in $^{132}$Sn.}
\label{figure9}
\end{figure}

\begin{figure}
\centering{\includegraphics[width=6cm]{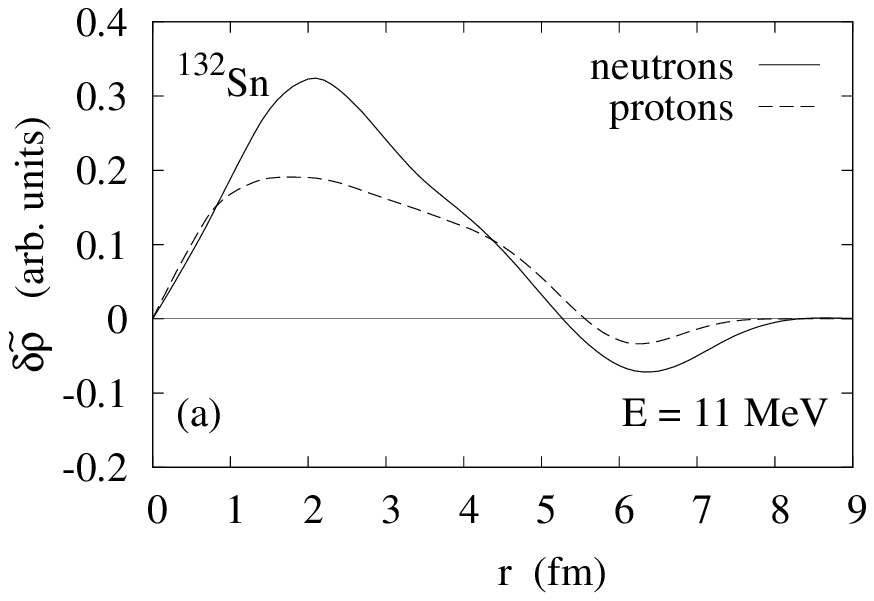}}\\
\centering{\includegraphics[width=6cm]{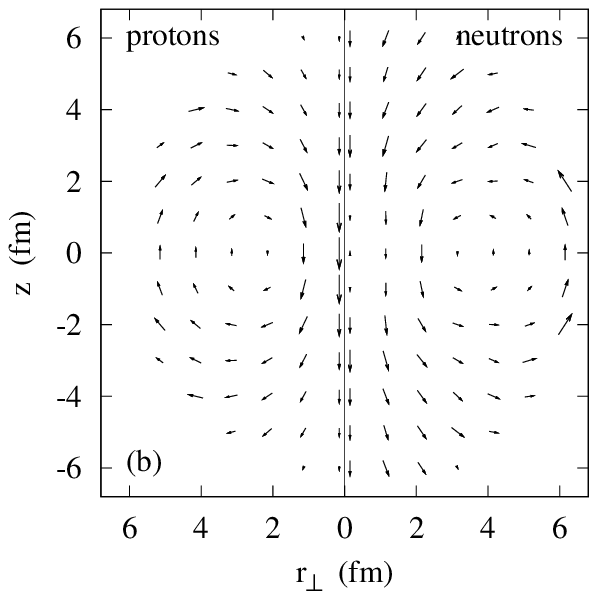}}
\caption{Same as \Fig{figure5}, but for the mode excited by the
toroidal dipole operator at $11$ MeV in $^{132}$Sn.}
\label{figure10}
\end{figure}

\begin{figure}
\centering{\includegraphics[width=6cm]{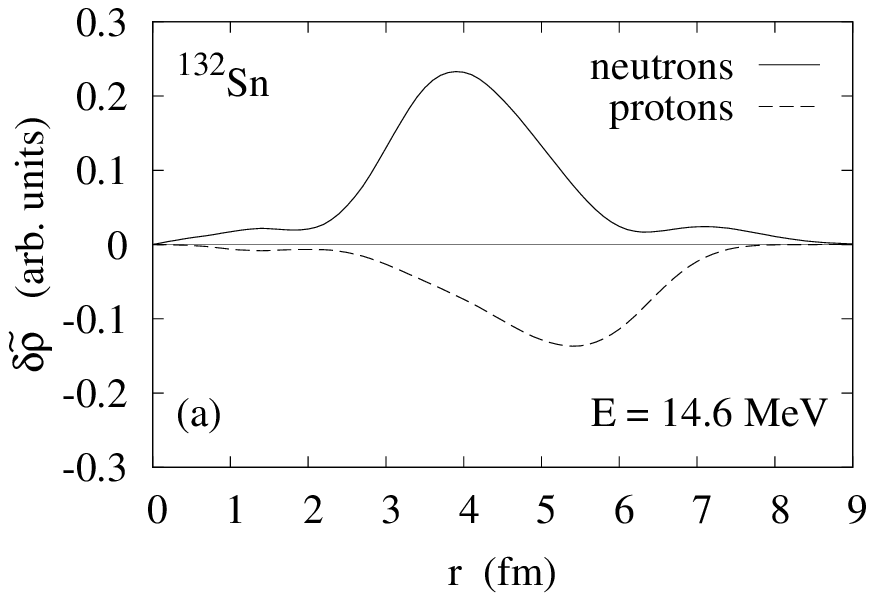}}\\
\centering{\includegraphics[width=6cm]{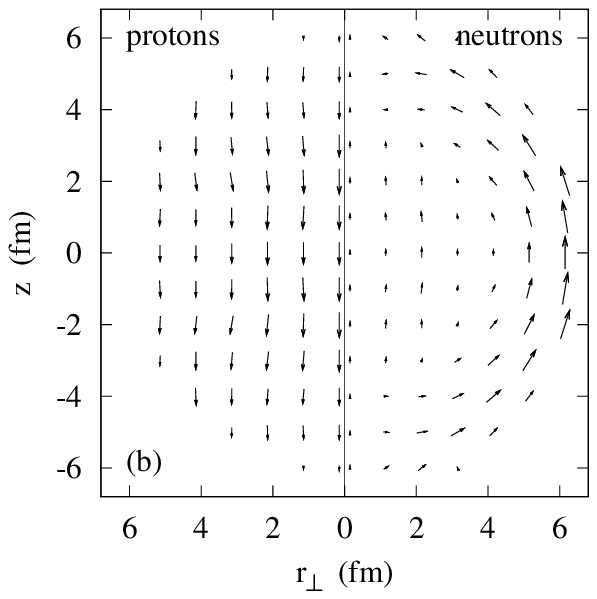}}
\caption{Same as \Fig{figure5}, but for the mode excited by the
toroidal dipole operator at $14.6$ MeV in $^{132}$Sn.}
\label{figure11}
\end{figure}

\begin{figure}
\centering{\includegraphics[width=6cm]{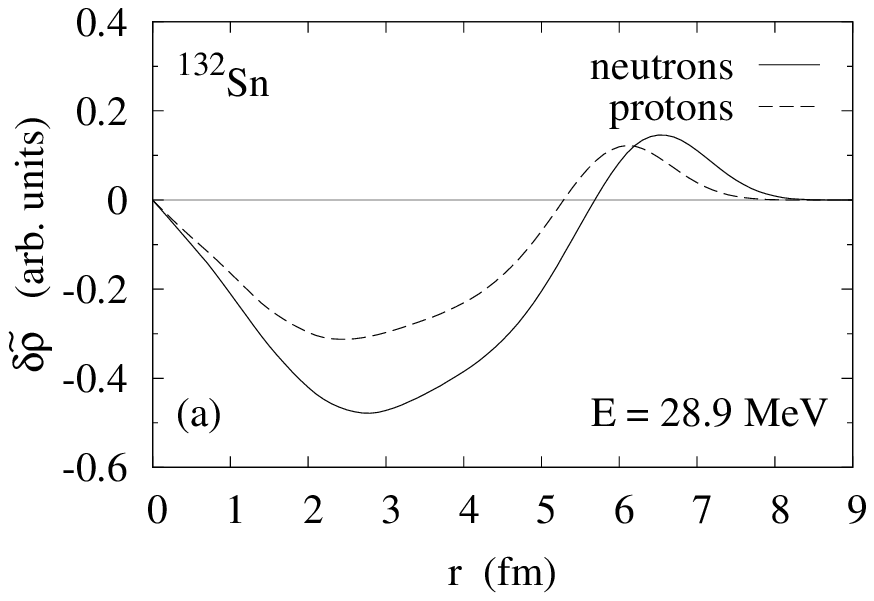}}\\
\centering{\includegraphics[width=6cm]{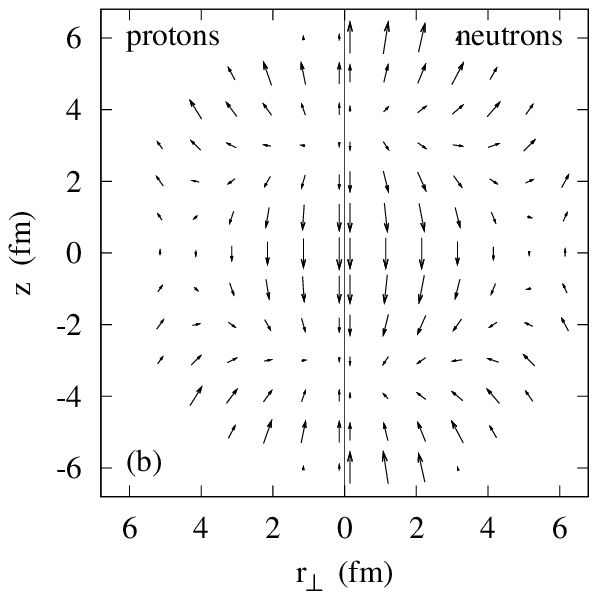}}
\caption{Same as \Fig{figure5}, but for the compressional dipole mode
at $28.9$ MeV in $^{132}$Sn.}
\label{figure12}
\end{figure}

In \Figs{figure9}-\ref{figure12} the transition densities and velocity
fields corresponding to the peaks of the toroidal dipole response of
$^{132}$Sn at $8.9$, $11$, $14.6$, and $28.9$ MeV are shown.
Let us first compare the results for the modes at $8.9$
(\Fig{figure9}) and $11$ MeV (\Fig{figure10}). At first glance, the
velocity fields look similar for these two modes, but the transition
densities are completely different. Both modes are essentially
isoscalar. Comparing the results for the $8.9$ MeV mode in
\Fig{figure9} with the results for the PDR at $8.6$ MeV in
\Fig{figure7}, we see a striking similarity. One can say that these
two modes are in fact one and the same, only excited in two different
ways. Note that the lines with zero velocity, around which the protons
and neutrons circulate, are at $\sim 3$ fm and $\sim 4$ fm from the
center of the nucleus, respectively, i.e., they lie inside the core of
the nucleus and not in the surface. The velocity field of the mode at
$11$ MeV looks even more like a torus, since the shape of the velocity
field is more rounded than in the mode at $8.9$ MeV. What about the
third mode at $14.6$ MeV (\Fig{figure11})? First of all, from the
transition densities, one sees that this mode has mainly isovector
character. Comparing with the results for the GDR at $14.5$ MeV in
\Fig{figure6}, one concludes that the mode at 14.6 MeV is in fact the
GDR, which is excited by the isoscalar toroidal dipole operator due to
the strong neutron excess in $^{132}$Sn. Finally, the high-lying
isoscalar dipole mode at $28.9$ MeV (\Fig{figure12}) has a completely
different nature. As can be seen from the velocity field, this mainly
isoscalar mode exhibits a compressional motion, and for this reason it
is usually called the compressional dipole mode.

\begin{figure}
\centering{\includegraphics[width=6cm]{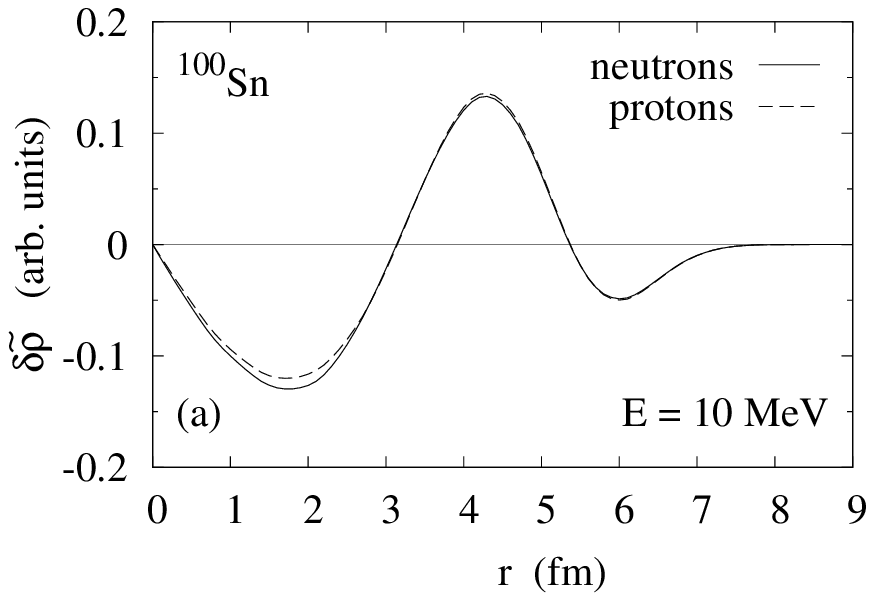}}\\
\centering{\includegraphics[width=6cm]{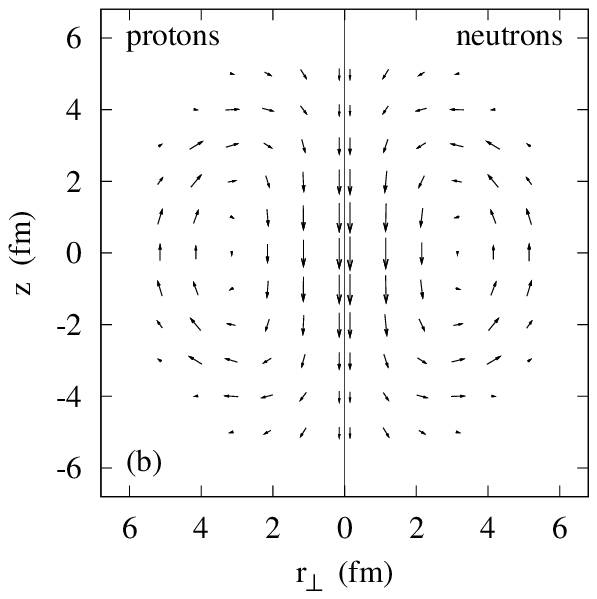}}
\caption{Same as \Fig{figure5}, but for the toroidal mode
at $10$ MeV in $^{100}$Sn.}
\label{figure13}
\end{figure}

\begin{figure}
\centering{\includegraphics[width=6cm]{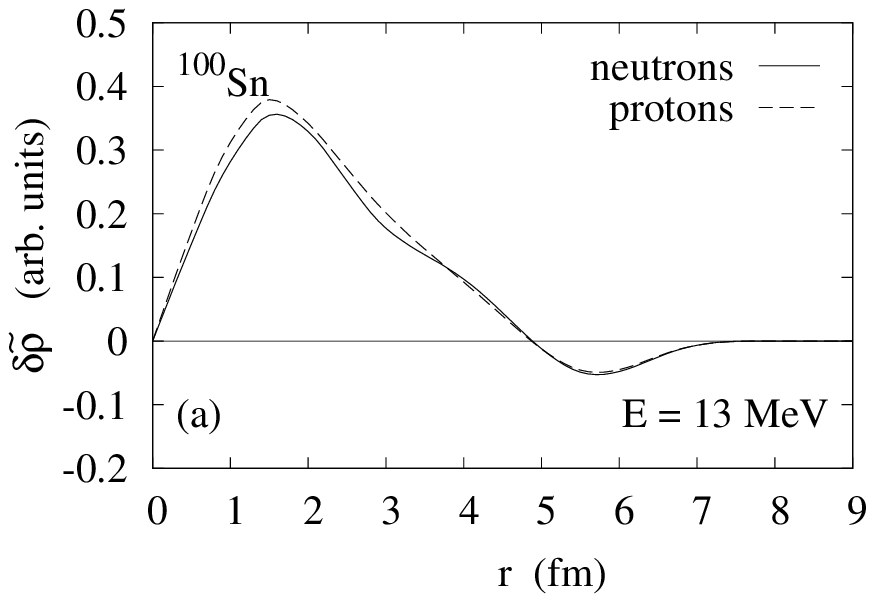}}\\
\centering{\includegraphics[width=6cm]{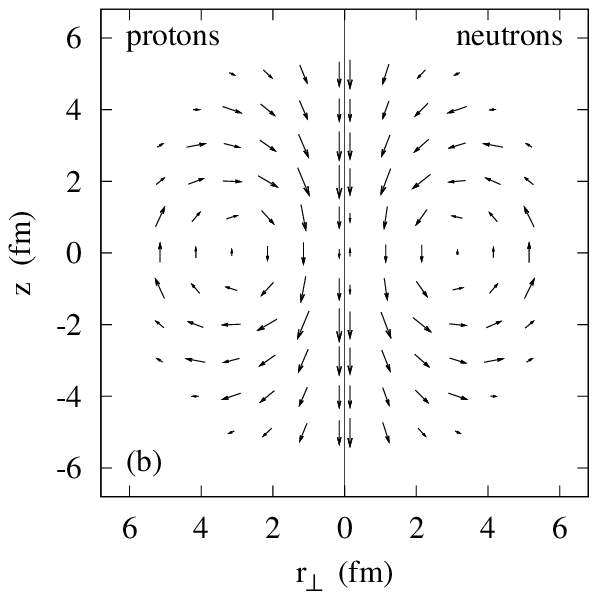}}
\caption{Same as \Fig{figure5}, but for the toroidal mode
at $13$ MeV in $^{100}$Sn.}
\label{figure14}
\end{figure}

In order to get a better understanding of the two low-lying modes at
$8.9$ and $11$ MeV, let us look at the corresponding modes of
$^{100}$Sn which lie at $10$ and $13$ MeV, see \Figs{figure13} and
\ref{figure14}.
We observe that the mode at $10$ MeV in $^{100}$Sn (\Fig{figure13})
has qualitatively the same velocity field and transition density as
the mode corresponding to the PDR in $^{132}$Sn (\Fig{figure9}). From
this one may conclude that the existence of this mode does not require
the presence of a neutron skin. The neutron excess is only needed in
order to be able to probe this mode with the electric dipole
operator. Looking closely at the velocity field of this mode
(\Figs{figure9} and \ref{figure13}), one realizes that the velocity
field is almost constant in the center of the nucleus, contrary to the
velocity field of the toroidal mode which lies at slightly higher
energy [$11$ MeV in $^{132}$Sn (\Fig{figure10}) and $13$ MeV in
  $^{100}$Sn (\Fig{figure14}), respectively]. This can be interpreted
in the sense hat the lower one of the two modes is an oscillation of
the surface (not necessarily the neutron skin, since $^{100}$Sn does
not have one) against the core, whereas the higher one is the original
torus mode which, in the framework of nuclear fluid dynamics, exists
already in a uniform sphere. This interpretation is corroborated by
the transition densities, which in the case of the higher mode
(\Figs{figure10} and \ref{figure14}) are much more concentrated in the
inner part of the nucleus, while those of the lower mode
(\Figs{figure9} and \ref{figure13}) are much stronger in the surface
region. Another support for this interpretation is the energy of this
mode: According to \Ref{Bastrukov1993}, in a uniform sphere and within
nuclear fluid dynamics, the torus mode should lie at $65-85 A^{-1/3}$
MeV, i.e., at $14-18.3$ MeV in the case of $^{100}$Sn and $12.8-16.7$
MeV in the case of $^{132}$Sn. This is slightly higher, but not very
far from the modes found here at $13$ and $11$ MeV, respectively.

\section{Summary and Discussion}
\label{sec:summary}
In this paper, electric and isoscalar dipole excitations were studied
within the semiclassical TF plus Vlasov approach. As interaction, the
bulk part of the BCP functional was employed, which was smoothed out
in space in order to mimic the effect of the neglected finite-range
term. The TF equation for the ground state as well as the Vlasov
equation for the dynamics were solved numerically without any further
simplifying assumptions. Compared to fully quantum mechanical
Hartree-Fock plus RPA calculations, the present method is missing the
shell effects, what can be useful if one is interested in generic
properties and average trends.

The electric dipole response of oxygen and tin isotopes was discussed
(calculations for other nuclei like calcium and lead were performed
but not shown). In all cases, low-lying strength corresponding to the
PDR was found in the very neutron-rich isotopes, however in the
neutron-rich oxygen isotopes the strength was spread over a large
energy range, in accordance with the strong fragmentation of the
strength in quantum mechanical calculations \cite{VretenarNPA}, so
that one cannot speak of a collective mode in this case. This shows
that the existence of the PDR is a generic property of neutron-rich
nuclei and does not rely on a specific structure of single-particle
levels. In heavier nuclei, the PDR is found to be a collective
excitation. The obtained energies and transition probabilities are
roughly in agreement with experimental data (at least as well as it
can be expected in a theory without shell effects). The transition
densities and velocity fields of the pygmy mode were analysed and it
was found that the velocity field has a toroidal shape, in agreement
with the findings of earlier quantum mechanical calculations
\cite{VretenarNPA,Ryezayeva}. The vortex line lies inside the core of
the nucleus, and the torus can therefore be seen in both neutron and
proton velocity fields, which suggests that the popular picture of the
neutron skin oscillating against a static $N=Z$ core is
oversimplified.

In order to compare the PDR and the torus mode in detail, the response
to the isoscalar toroidal operator was studied. In the example of
$^{132}$Sn it was found that, due to the mixing of isoscalar and
isovector modes, the predominantly isoscalar PDR and the predominantly
isovector GDR can be seen in both the E1 and the toroidal strength
functions. In addition, the toroidal response exhibits two peaks which
do not show up in the electric dipole response: a second toroidal mode
which lies slightly above the PDR, and a compressional dipole
mode. The two toroidal modes exist also in the $N=Z$ nucleus
$^{100}$Sn. Hence, the existence of these modes, including the lower
one which in the case of $^{132}$Sn was identified with the PDR, does
not rely on the presence of a neutron skin. If this interpretation is
correct, the reason why the PDR is only seen in nuclei with large
neutron excess is simply that otherwise the E1 strength of this mode
is too small to be seen. However, it was recently pointed out that, if
the isospin symmetry breaking effect of the Coulomb interaction is
taken into account, a small contribution of this mode can be seen even
in the electric dipole response of $N=Z$ nuclei
\cite{Papakonstantinou}). Another conclusion which can be drawn from
this result is that the PDR, like the torus mode, cannot be described
in a hydrodynamical picture, but its existence relies on the
``elasticity'' of the nuclear medium due to Fermi-surface
deformation. This fact was also stressed in \Ref{Bastrukov2008}.

The two toroidal modes are apparently qualitatively different,
although their velocity fields look quite similar: The higher mode
corresponds to the torus mode of an elastic sphere, which has been
discussed in the literature for many years \cite{Semenko,Bastrukov1993},
whereas the lower one corresponds more to an oscillation of the core
against the surface (but not necessarily against the neutron skin),
qualitatively similar to the modes discussed in \Ref{Bastrukov2008}.

Of course, the present approach has some shortcomings and is not meant
to replace more sophisticated quantum-mechanical calculations. Since
it is a semiclassical formalism, shell effects cannot be described. On
the one hand, this results in clear pictures for the different modes,
but on the other hand, it is of course a simplification which makes
the detailed comparison with experiment and with more realistic
calculations difficult. In addition, a couple of approximations were
made which give probably rise to systematic deviations. For example,
the Coulomb interaction was omitted. One would also expect an
important effect from pairing, since it affects in particular
rotational motion and excitations involving Fermi surface
deformation. Nevertheless, the obtained results are surprisingly
reasonable and will maybe serve as a motivation for a more detailed
study of the relationship between the pygmy and the torus mode within
fully quantum mechanical approaches.

\begin{acknowledgments}
I thank E. Balbutsev and P. Papakonstantinou for discussions and
P. Schuck for numerous ideas, suggestions and careful reading of the
manuscript. This work was supported by ANR (project NEXEN).
\end{acknowledgments}

\appendix
\section{Angle averaged densities and velocity fields}
All simulations were done in three dimensions without any imposed
symmetries. However, for the graphical representation of the results
it is advantageous to average the densities and velocity fields over
the angle in order to reduce the statistical noise due to the finite
number of test particles.

Let us start with the ground state density distributions shown in
\Fig{figure1}. Within the TF approximation, the ground states of all
nuclei are spherical.  Therefore, the ground state densities
$\tilde{\rho}$ were averaged over the full solid angle. In terms of
the test-particle positions $\rv_i$, the angle-averaged densities can
be expressed as
\begin{equation}
\tilde{\rho}(r) = \sum_{i=1}^{\calN A} \frac{e^{-(r-r_i)^2/d^2}-e^{-(r+r_i)^2/d^2}}
  {\calN 4 \pi^{3/2} r r_i d}\,.
\end{equation}

The dipole excitations considered in this work destroy the spherical
symmetry, but not the cylindrical symmetry with respect to the $z$
axis.  In \Figs{figure5}-\ref{figure7} and
\ref{figure9}-\ref{figure14}, this symmetry was used to reduce the
statistical fluctuations by averaging the density $\tilde{\rho}$ and
the components $\tilde{j}_\perp$ and $\tilde{j}_z$ of the current
density over the azimuthal angle $\phi$ ($\tilde{j}_\phi = 0$ for the
excitation operators under consideration). After this averaging, the
final expressions for $\tilde{\rho}$ and $\tilde{\jv}$ in terms of the
test particle positions $\rv_i$ and momenta $\pv_i$ read:
\begin{gather}
\tilde{\rho}(r_\perp,z) = \sum_{i=1}^{\calN A}
  \frac{e^{-\frac{(z-z_i)^2+r_\perp^2+r_{\perp i}^2}{d^2}}}
    {\calN(\sqrt{\pi}d)^3} 
  I_0\Big(\frac{2r_\perp r_{\perp i}}{d^2}\Big)\,,\\
\tilde{j}_z(r_\perp,z) = \sum_{i=1}^{\calN A} p_{z i}
  \frac{e^{-\frac{(z-z_i)^2+r_\perp^2+r_{\perp i}^2}{d^2}}}
    {\calN(\sqrt{\pi}d)^3}
  I_0\Big(\frac{2r_\perp r_{\perp i}}{d^2}\Big)\,,\\
 \tilde{j}_\perp(r_\perp,z) 
  = \sum_{i=1}^{\calN A} p_{\perp i}\cos\phi_{r_ip_i}
  \frac{e^{-\frac{(z-z_i)^2+r_\perp^2+r_{\perp i}^2}{d^2}}}
    {\calN(\sqrt{\pi}d)^3} I_1\Big(\frac{2r_\perp r_{\perp i}}{d^2}\Big)\,,
\end{gather}
where $I_0$ and $I_1$ are modified Bessel functions \cite{Abramowitz}
and $\phi_{r_ip_i}$ is the difference of the azimuthal angles of
$\rv_i$ and $\pv_i$, i.e.,
\begin{equation}
p_{\perp i}\cos\phi_{r_ip_i} = \frac{x_i p_{xi}+y_i p_{yi}}
  {\sqrt{x_i^2+y_i^2}}\,.
\end{equation}
\section{Transition probabilities and energy weighted sum rule}
Often, the $B(E1)$ value of the pygmy mode or its contribution to the
EWSR are used as a measure for the strength of the pygmy
mode. According to the definitions given in \Ref{BohrMottelson}, the
reduced transition probability $B(E1;0\to 1)$ from the $I_1=0$ ground
state to a $I_2=1$ excited state can be related to the strength
function $S(E)$ corresponding to the electric dipole operator
(\ref{eldipole}) as follows:
\begin{equation}
\frac{dB(E1;0\to 1)}{dE} = \frac{9e^2}{4\pi} S(E)\,.
\end{equation}
Since the BCP functional does not introduce an effective mass ($m^* =
m$), the EWSR (Thomas-Reiche-Kuhn sum rule)
\begin{equation}
\int_0^\infty dE E S(E) = \frac{\hbar^2}{2m} \frac{NZ}{A}\,.
\end{equation}
should be exactly fulfilled. In the numerical results discussed in
\Sec{subsec:electric}, the deviation from the exact result is less
than 1\%.


\end{document}